\documentclass[aps,superscriptaddress,prd,nofootinbib,10pt]{revtex4-2} %
\usepackage{graphicx}%
\usepackage[caption=false,singlelinecheck=false]{subfig}%
\usepackage{amsmath,amssymb,amsfonts,amsthm,mathrsfs}%
\usepackage{booktabs,txfonts,paralist,enumitem,ulem}%
\usepackage{tikz}%
\usepackage[colorlinks,urlcolor=black,citecolor=black,link
color=black]{hyperref}

\newcommand{\red}[1]{\color{red} #1 \color{black}}
\newcommand{\blue}[1]{{\color{blue} #1\color{black}}}

\allowdisplaybreaks

\begin{document}

\title{Comments on ``Exploring Quantum Statistics for Dirac and
Majorana Neutrinos using Spinor-Helicity techniques''
(arXiv:2507.07180 [hep-ph])}

\author{C.~S.~Kim}%
\email[Email at: ]{cskim@yonsei.ac.kr}%
\affiliation{Department of Physics and IPAP, Yonsei University, Seoul
03722, Korea}%

\author{M.~V.~N.~Murthy}%
\email[Email at: ]{murthy@imsc.res.in}%
\affiliation{The Institute of Mathematical Sciences, Taramani, Chennai
600113, India}%

\author{Dibyakrupa~Sahoo}%
\email[Email at: ]{Dibyakrupa.Sahoo@fuw.edu.pl}%
\affiliation{Faculty of Physics, University of Warsaw, Pasteura 5,
02-093 Warsaw, Poland}

\date{\today}

\begin{abstract}
We give our comments on Ref.~\cite{Bigaran:2025kod} which critiques the idea of using  quantum statistics to distinguish between Dirac and Majorana neutrinos proposed in some of our earlier works \cite{Kim:2021dyj, Kim:2023iwz, Kim:2024tsm}. The ad hoc symmetrization of the Dirac case amplitude square advocated in Eqs.~(16) and (35) of \cite{Bigaran:2025kod} has no physical basis and it leads to violation of lepton number in the standard model for Dirac neutrinos. Therefore, this symmetrization `by hand' is in principle incorrect.
\end{abstract}
\maketitle


\section{Note Added after publication of [1]}\label{sec:note-added}

In the published version of the paper ~\cite{Bigaran:2025kod} we find essentially the same mistakes which we had pointed out before (see Sec.~\ref{sec:original-comments} for a concise version of our original comments).

\begin{enumerate}
\item  Their Dirac case amplitude square for the doubly weak charged decay of the $B^0$ meson: $B^0 \to \mu^-(k) \bar\nu(p_1) \mu^+(\bar{k}) \nu(p_2)$, as given in their Eq.~(35) has two terms. The second term has been added by hand to their Eq.~(34). The first and second terms of their Eq.~(35) are related to each other via exchange of the neutrino antineutrino
4-momenta, $p_1 \leftrightarrow p_2$. This makes their Dirac case
amplitude square symmetric under $p_1 \leftrightarrow p_2$, even
though Dirac neutrino and antineutrino are distinguishable particles.
The addition of the $p_1 \leftrightarrow p_2$ exchanged second term
without simultaneously interchanging the muon 4-momenta, ie. $k
\leftrightarrow \bar{k}$, implies that the lepton number is violated
for the second term of their Eq.~(35). 
The anti-symmetrization of amplitude for Majorana neutrino is
enforced by Pauli principle due to the identical nature of Majorana
neutrino and antineutrino, and it is unrelated to detection or non-detection of the final neutrinos. Therefore, the ad hoc symmetrization of
Dirac case amplitude square under $p_1 \leftrightarrow p_2$ exchange
in Eq.~(35) of \cite{Bigaran:2025kod} is incorrect.

\item The Eqs.~(16) and (35) of \cite{Bigaran:2025kod} seem to suggest
that if one does not observe some of the non-identical final state 
particles, then one needs to symmetrize the amplitude square of 
the process by adding terms that exchange the unobserved 4-momenta.
We find this as a rather strange claim. It is a well known fact that 
in any collider experiment some particles escape detection or 
identification, when they are either very close to the beam pipe or 
when the tracks get too crowded or if a part of the detector 
malfunctions or if many long-lived particles are 
present in the final state which decay outside the detector, etc.
In such cases the amplitude square of the process is never symmetrized 
for non-detection of any of the distinct particles. Instead, the 
amplitude and amplitude square of the process are calculated 
assuming that all the non-identical particles are detectable 
and distinguishable in principle, and the non-detection of any 
particle is taken care of by appropriately constraining the 
phase space integral.

\item In our paper \cite{Kim:2021dyj}, we have used a simplified 
notation $M(p_1, p_2)$ to denote the amplitude of the doubly weak charged decay of the 
$B^0$ meson: $B^0 \to \mu^-(p_-) \bar\nu(p_1) \mu^+(p_+) \nu(p_2)$. 
Our simplified notation for the Dirac case amplitude, as given in 
Eq.~(2) of our paper \cite{Kim:2021dyj} has the following expanded 
form, 
$$M^D = M(p_1, p_2) \equiv M(p_{-}, s_{-}; p_{1}, s_{1}; p_{+}, s_{+}; p_2, s_2)$$ 
where we have now explicitly mentioned all the 4-momenta and spins of
the final particles. For brevity of expressions we chose to highlight
only the $p_1,p_2$ dependence, while other dependencies were
implicitly always there. As mentioned below our Eq.~(3) in
\cite{Kim:2021dyj}, the amplitude squares, e.g.\ $|M^D|^2$, include
summation over all final spins and averaging of initial spins. Note
that since the parent particle $B^0$ has spin $0$, the average over
initial spins is trivial/meaningless in our case.

\item Just above Eq. (43) of \cite{Bigaran:2025kod}, the authors 
claimed that  ``... exactly vanishes when $p_1=p_2$ ... ."  This 
sentence is absolutely misleading. As explained above, the spin 
information is implicit in our simplified amplitude notation. 
Therefore, the correct statement would have been ``... exactly 
vanishes when $p_1=p_2$ and $s_1=s_2$ ... .''
\end{enumerate}

\section{Summary of our Original Comments}\label{sec:original-comments}

In Ref.~\cite{Bigaran:2025kod} the authors aim to provide a critique
of our papers \cite{Kim:2021dyj, Kim:2023iwz, Kim:2024tsm} where we
have explored the possibility whether quantum statistics could help
distinguish Dirac and Majorana neutrinos. We note that in
Ref.~\cite{Bigaran:2025kod} the authors have tried to force agreement
with the practical Dirac Majorana confusion theorem (pDMCT), while in
our works \cite{Kim:2021dyj, Kim:2023iwz, Kim:2024tsm} we have
explored the domain of applicability of pDMCT in addition to providing
a general proof of it in context of final states having neutrino
antineutrino pair. We have used the standard quantum field theory in
our papers \cite{Kim:2021dyj, Kim:2023iwz, Kim:2024tsm} and our
approach very well accommodates the previously existing literature.
Our comments are as follows.

\begin{enumerate}[leftmargin=0pt]
\item \red{ The underlying basis of the main claim of
Ref.~\cite{Bigaran:2025kod} where they disagree with our papers, is
contained within their Eqs.~(33)-(38) of Sec.~IV, as well as in
Eqs.~(14)-(16) of their Sec.~III. While discussing the case of Dirac
neutrinos, the authors of \cite{Bigaran:2025kod} contend that for the
unobserved final states of neutrino and antineutrino one should
symmetrize the amplitude squared with respect to the 4-momenta of the
neutrino and antineutrino. Just below Eq.~(34) of
\cite{Bigaran:2025kod} they state that ``To make a fair comparison
with the Majorana case, as the neutrinos are not observed, it is
imperative that \textit{both} momentum assignments be the
anti-neutrino are summed to construct the Dirac matrix-element
squared''. } This notion of symmetrizing at the probability level for
Dirac neutrinos affects the inferences made in
Ref.~\cite{Bigaran:2025kod}.

\begin{enumerate}
\item \blue{There is no fundamental principle or law of physics which
requires that the amplitude square for a process having two
distinguishable particles in the final state ought to be symmetrized
with respect to the exchange of the 4-momenta of the concerned
particles, when the two particles are not detected in the
detector}\footnote{We note that with advancement of technology,
particle detection and identification also improves, and even with the
best of detector technology some particles might not get detected by
the detector either due to limited spatial coverage of the detector or
due to the finite time needed for triggering events in a detector.}. A
Dirac neutrino is distinguishable from the corresponding antineutrino.
Therefore, for a final state containing a pair of Dirac neutrino and
antineutrino there is no need to do any symmetrization with respect to
their momentum exchange.

\blue{Detection and identification of a particle in a detector does
not affect the amplitude square which is a purely theoretical
computation where one assumes that all the 4-momenta of all the
particles are well known.} The quantity which gets affected by
experimental detection or observation or measurement is the
``observable'' (such as some asymmetry, or distribution, or
differential decay rate, or scattering cross-section etc.) which
involves both the amplitude square as well as the phase space
considerations. Non-observation of any final particle is well taken
care of by suitably performing corresponding phase space integration.
We have already pointed out in the beginning of Sec.~III of our paper
\cite{Kim:2021dyj} as well as in Sec.2.2 of \cite{Kim:2023iwz} that
such phase space integration for invisible neutrinos directly leads to
pDMCT, and this does not require any ad hoc symmetrization of the
Dirac case amplitude square as done in \cite{Bigaran:2025kod}.

\item \blue{ In the standard model (SM) the Dirac neutrino ($\nu$) and
antineutrino ($\overline{\nu}$) carry lepton numbers as they are
produced via different weak charged currents (CCs) $W^+ \to \ell^+
\nu_\ell$ and $W^- \to \ell^- \overline{\nu}_\ell$ for $\ell
=e,\mu,\tau$, to ensure lepton number conservation within the SM. }
When considering Dirac neutrino antineutrino pair production via two
CCs, as it happens in the decays considered in
Ref.~\cite{Bigaran:2025kod} as well as in our work \cite{Kim:2021dyj},
the neutrino and antineutrino 4-momenta are paired with the
corresponding charged lepton so that lepton number is conserved.
\blue{ This is not required for Majorana neutrinos, for which
$\nu=\overline{\nu}$ and the lepton number is not a good (conserved)
quantum number. However, adding a term to the Dirac case amplitude
square where the neutrino and antineutrino 4-momenta have been
exchanged, as done in the right sides of Eqs.~(16) and (35) of
Ref.~\cite{Bigaran:2025kod}, amounts to considering $W^+ \to \ell^+
\overline{\nu}_{\ell}~(\Delta L=-2)$ and $W^- \to \ell^-
\nu_{\ell}~(\Delta L=+2)$ both of which violate lepton number. } When
working with Dirac neutrinos in the framework of the SM alone without
consideration of any new physics contributions, as is the case in
decay $B^0 \to \mu^- \mu^+ \nu_\mu \overline{\nu}_\mu$ considered in
both \cite{Bigaran:2025kod} and \cite{Kim:2021dyj}, one should avoid
considering lepton number violation effects.

\end{enumerate}

\item \red{ The authors of Ref.~\cite{Bigaran:2025kod} state in the
paragraph containing Eq.~(32) that we have not been explicit about how
we construct the Dirac as well as Majorana amplitudes in
\cite{Kim:2021dyj}. } \blue{ We note that we have explicitly and fully
mentioned construction of our Dirac case and Majorana case amplitudes
in Sec.~IV, sub-sections A, B, C and D of \cite{Kim:2021dyj}. As far
as the spin or helicity of the neutrinos are concerned, they are
automatically taken care of by the $V-A$ nature of the CC vertices. }
To elaborate slightly more on this let us first focus on the Dirac
case, the 4-momenta of antineutrino and neutrino in \cite{Kim:2021dyj}
are denoted by $p_1$ and $p_2$ respectively. The amplitude in our case
is denoted by $\mathscr{M}^D \equiv \mathscr{M}(p_1,p_2)$ and given by
Eq.~(15a) in \cite{Kim:2021dyj}.\footnote{The momentum-exchanged
amplitude $\mathscr{M}(p_2,p_1)$, is not at all required in the Dirac
case. There is \textit{no underlying symmetry principle} that connects
the two amplitudes, $\mathscr{M}(p_1,p_2)$ and $\mathscr{M}(p_2,p_1)$,
\textit{in the general case}.} We then follow the \textit{standard
procedure} in any QFT calculation to evaluate the amplitude square in
such a case, i.e.\ we take the modulus square of the amplitude, sum
over final spins, average over initial spins\footnote{Since $B^0$ is a
pseudo-scalar meson its spin is $0$ and no averaging over initial
spins is needed.}, and evaluate the traces that appear during the
calculation. We clarify that our calculation in \cite{Kim:2021dyj}
does not explicitly consider specific helicities of the neutrino and
antineutrino.

\blue{ We again reiterate that \textit{the problem of detectability of
particles is still not relevant at the amplitude stage}. As mentioned
before, it is relevant only when computing any observable for
experimental study, in which case we need to integrate over all the
unobservable phase space variables so that the final observable thus
obtained involves only measurable quantities. We emphasize, there is
no formal reasoning for symmetrizing the amplitude squared as done in
Ref.~\cite{Bigaran:2025kod}. } Even when required by quantum
statistics, any symmetrization (for identical bosons) or
antisymmetrization (for identical fermions) is done at the level of
{\it amplitude only}. The question of detectibility can neither alter
the rules for writing the transition amplitude nor the computation of
the amplitude square.

\item \red{ The authors of Ref.~\cite{Bigaran:2025kod} wrongly presume
that our calculation in \cite{Kim:2021dyj} follows steps analogous to
what they suggest in their Eqs.~(40) and (43). } As clarified above,
the calculation in \cite{Kim:2021dyj} does not explicitly consider any
specific helicity states for the neutrino and antineutrino pair in the
final state. The suggestion made in Eq.~(28) of
Ref.~\cite{Kim:2023iwz} is a generic suggestion where the decay
amplitude can be written as a sum of a set of mutually orthogonal
helicity amplitudes. \blue{ Although we have not used the helicity
amplitude formalism, as mentioned before and as we have noted below
Eq.~(30c) of \cite{Kim:2021dyj}, we have summed over all final spins
in our calculations  instead. }
\end{enumerate}

We have considered the claims made in Ref.~\cite{Bigaran:2025kod} in
the context of our papers \cite{Kim:2021dyj, Kim:2023iwz, Kim:2024tsm}
and presented clarifications to consider them both invalid and
incorrect. \blue{ The main crux of the arguments presented in our
papers~\cite{Kim:2021dyj, Kim:2023iwz, Kim:2024tsm} is to show that
quantum statistics can not only distinguish between Majorana or Dirac
types, but also define the domain of validity of pDMCT in the most
general way possible in the absence of a formal proof of it until
now.} We note that pDMCT is strictly valid when we integrate over the
momenta of neutrino and antineutrino (using the Dirac nomenclature).
The proposed ad hoc symmetrisation in \cite{Bigaran:2025kod} at the
probability level for the Dirac case simply achieves this by a
different procedure, but as explained before, we do not consider this
approach to be correct. Nevertheless, our main focus in
\cite{Kim:2021dyj, Kim:2023iwz, Kim:2024tsm, Kim:2022xjg,
Hernandez-Tome:2024oxb} is on the possible exceptions to pDMCT when we
can fix the neutrino-antineutrino momenta either directly (possible in
principle but difficult experimentally) or indirectly (as pointed out
in \cite{Kim:2021dyj} under special kinematic conditions) or when new
physics effects are taken into consideration.

\acknowledgments

The work of CSK is supported by Basic Science Research Program through the National Research Foundation of Korea (NRF) funded by the Ministry of Education (RS-2022-NR070836, RS-2026-25471898).
We thank the authors of \cite{Bigaran:2025kod},
especially Innes Bigaran and Stephen Parke, for engaging in discussions on our paper
that brings clarity to many of the issues.


\begin{thebibliography}{99}

\bibitem{Bigaran:2025kod}
I.~Bigaran, S.~J.~Parke and P.~Pasquini,
``Exploring Quantum Statistics for Dirac and Majorana Neutrinos using Spinor-Helicity techniques,''
Phys. Rev. D \textbf{112}, no.5, 053008 (2025)
[arXiv:2507.07180 [hep-ph]].


\bibitem{Kim:2021dyj}
C.~S.~Kim, M.~V.~N.~Murthy and D.~Sahoo,
``Inferring the nature of active neutrinos: Dirac or Majorana?,''
Phys. Rev. D \textbf{105}, no.11, 113006 (2022)
doi:10.1103/PhysRevD.105.113006
[arXiv:2106.11785 [hep-ph]].


\bibitem{Kim:2023iwz}
C.~S.~Kim,
``Practical Dirac Majorana confusion theorem: issues and applicability,''
Eur. Phys. J. C \textbf{83}, no.10, 972 (2023)
doi:10.1140/epjc/s10052-023-12156-9
[arXiv:2307.05654 [hep-ph]].


\bibitem{Kim:2024tsm}
C.~S.~Kim, D.~Sahoo and K.~N.~Vishnudath,
``Searching for signatures of new physics in $B \rightarrow K \, \nu \, \overline{\nu }$ to distinguish between Dirac and Majorana neutrinos,''
Eur. Phys. J. C \textbf{84}, no.9, 882 (2024)
doi:10.1140/epjc/s10052-024-13262-y
[arXiv:2405.17341 [hep-ph]].


\bibitem{Kim:2022xjg}
C.~S.~Kim, J.~Rosiek and D.~Sahoo,
``Probing the non-standard neutrino interactions using quantum statistics,''
Eur. Phys. J. C \textbf{83}, no.3, 221 (2023)
doi:10.1140/epjc/s10052-023-11355-8
[arXiv:2209.10110 [hep-ph]].

\bibitem{Hernandez-Tome:2024oxb}
G.~Hern{\'a}ndez-Tom{\'e}, C.~S.~Kim and G.~L.~Castro,
Eur. Phys. J. C \textbf{85}, no.6, 686 (2025)
doi:10.1140/epjc/s10052-025-14379-4
[arXiv:2411.09124 [hep-ph]].

\end{thebibliography}
\end{document}